\documentclass[
preprintnumbers,reprint,
amssymb,notitlepage,
aps,prl,superscriptaddress
]{revtex4-1}
\usepackage{amsmath}
\usepackage{graphicx}
\usepackage{dcolumn}
\usepackage{bm}
\usepackage{braket}
\usepackage{color}
\usepackage{enumitem}
\usepackage{hyperref}
\hypersetup{
	colorlinks=true,
	linkcolor=blue,
	filecolor=blue,
	citecolor=blue,  
	urlcolor=black,
}

\newtheorem{theorem}{Theorem}

\newtheorem{lemma}[theorem]{Lemma}

\newenvironment{proof}[1][Proof]{\noindent\textbf{#1.} }{\ \rule{0.5em}{0.5em}}
\def\be{\begin{equation}}
	\def\ee{\end{equation}}
\def\ba{\begin{eqnarray}}
	\def\ea{\end{eqnarray}}

\begin{document}
\preprint{MIT-CTP/4895}
\title{Superadditivity of the Classical Capacity with Limited Entanglement Assistance}
\author{Elton Yechao Zhu}
\email{eltonzhu@mit.edu}
\affiliation{
Department of Physics, 
Massachusetts Institute of Technology, Cambridge, Massachusetts 02139, USA}
\affiliation{Center For Theoretical Physics,
Massachusetts Institute of Technology, Cambridge, Massachusetts 02139, USA}
\author{Quntao Zhuang}%
\affiliation{
Department of Physics, 
Massachusetts Institute of Technology, Cambridge, Massachusetts 02139, USA}
\affiliation{Research Laboratory of Electronics, 
Massachusetts Institute of Technology, Cambridge, Massachusetts 02139, USA}
\author{Peter W. Shor}
\affiliation{Center For Theoretical Physics,
Massachusetts Institute of Technology, Cambridge, Massachusetts 02139, USA}
\affiliation{
Department of Mathematics, 
Massachusetts Institute of Technology, Cambridge, Massachusetts 02139, USA}
\date{\today}
	
\begin{abstract}
Finding the optimal encoding strategies can be challenging for communication using quantum channels, as classical and quantum capacities may be superadditive. Entanglement assistance can often simplify this task, as the entanglement-assisted classical capacity for any channel is additive, making entanglement across channel uses unnecessary. If the entanglement assistance is limited, the picture is much more unclear. Suppose the classical capacity is superadditive, then the classical capacity with limited entanglement assistance could retain superadditivity by continuity arguments. If the classical capacity is additive, it is unknown if superadditivity can still be developed with limited entanglement assistance. We show this is possible, by providing an example. We construct a channel for which the classical capacity is additive, but that with limited entanglement assistance can be superadditive. This shows entanglement plays a weird role in communication, and we still understand very little about it. 
\end{abstract}
\pacs{Valid PACS appear here}
\maketitle
In Shannon's classical information theory \cite{Shannon48}, a classical (memoryless) channel is a probabilistic map from input states to output states. This has been extended to the quantum world. A (memoryless) quantum channel is a time-invariant completely positive trace preserving (CPTP) linear map from input quantum states to output quantum states~\cite{Choi75}. A classical channel can only transmit classical information, and the maximum communication rate is fully characterized by its capacity. A quantum channel can be used to transmit not only classical information but also quantum information. Hence, there are different types of capacity, such as classical capacity $\mathcal{C}$ for classical communication~\cite{Holevo98,Schumacher97} and quantum capacity $\mathcal{Q}$ for quantum communication~\cite{Lloyd97,Shor02,Devetak05}.  
	
Since quantum channels transmit quantum states, and quantum states can be entangled with other parties, it is natural to ask if entanglement can assist the communication. This was first considered by Bennett \textit{et al.}, who showed that unlimited preshared entanglement could improve the classical capacity of a noisy channel~\cite{Bennett99,Bennett02}. Shor examined the case where only finite preshared entanglement is available and obtained a trade-off curve that illustrates how the optimal rate of classical communication depends on the amount of entanglement assistance (CE trade-off)~\cite{Shor04}. One can also consider how entanglement (E), classical communication (C), or quantum communication (Q) can trade-off against each other as resources. The tradeoff capacity of almost any two resources was studied by Devetak \textit{et al.}~\cite{DevetakShor05,Devetak08}, such as entanglement-assisted quantum capacity (QE trade-off). Subsequently, the triple resource (CQE) trade-off capacity was also characterized~\cite{Hsieh10,HsiehWilde10,Wilde12}.
	
However, almost all the capacity formulas above are given by regularized expressions. They are difficult to evaluate because they require an optimization over an infinite number of channel uses, which is typically intractable. The existence of this regularization is because entanglement across different channel uses can sometimes protect information against noise and improve the communication rate, a phenomenon often called superadditivity. Superadditivity has long been known to be the case for quantum capacity \cite{Divincenzo98,Smith07}, but remained undiscovered for classical capacity until Hastings gave an example~\cite{Hastings09}. One exception is the entanglement-assisted classical capacity $ \mathcal{C}_E $~\cite{Bennett02,Note1}. An intuitive understanding of the additivity of $ \mathcal{C}_E $ is that the best way to use entanglement is to preshare it to the receiver, but not across different channels.
	
The need for regularization for various capacity formulas represents our incomplete understanding of quantum channels, as one cannot find the optimal transmission rate and best encoding strategies. Thus, an important goal in quantum Shannon theory is to characterize quantum channels with additive capacities. For classical capacity, many such channels are known, including unital qubit channels~\cite{King02}, entanglement-breaking channels~\cite{Shor02EB}, etc. For quantum capacity, there are also examples like degradable channels \cite{DevetakShor05}. Additivity for the double or triple resource trade-off capacity has also been considered, but many fewer examples are known~\cite{Bradler10}.
	
\begin{figure}
\includegraphics[width=0.45\textwidth]
{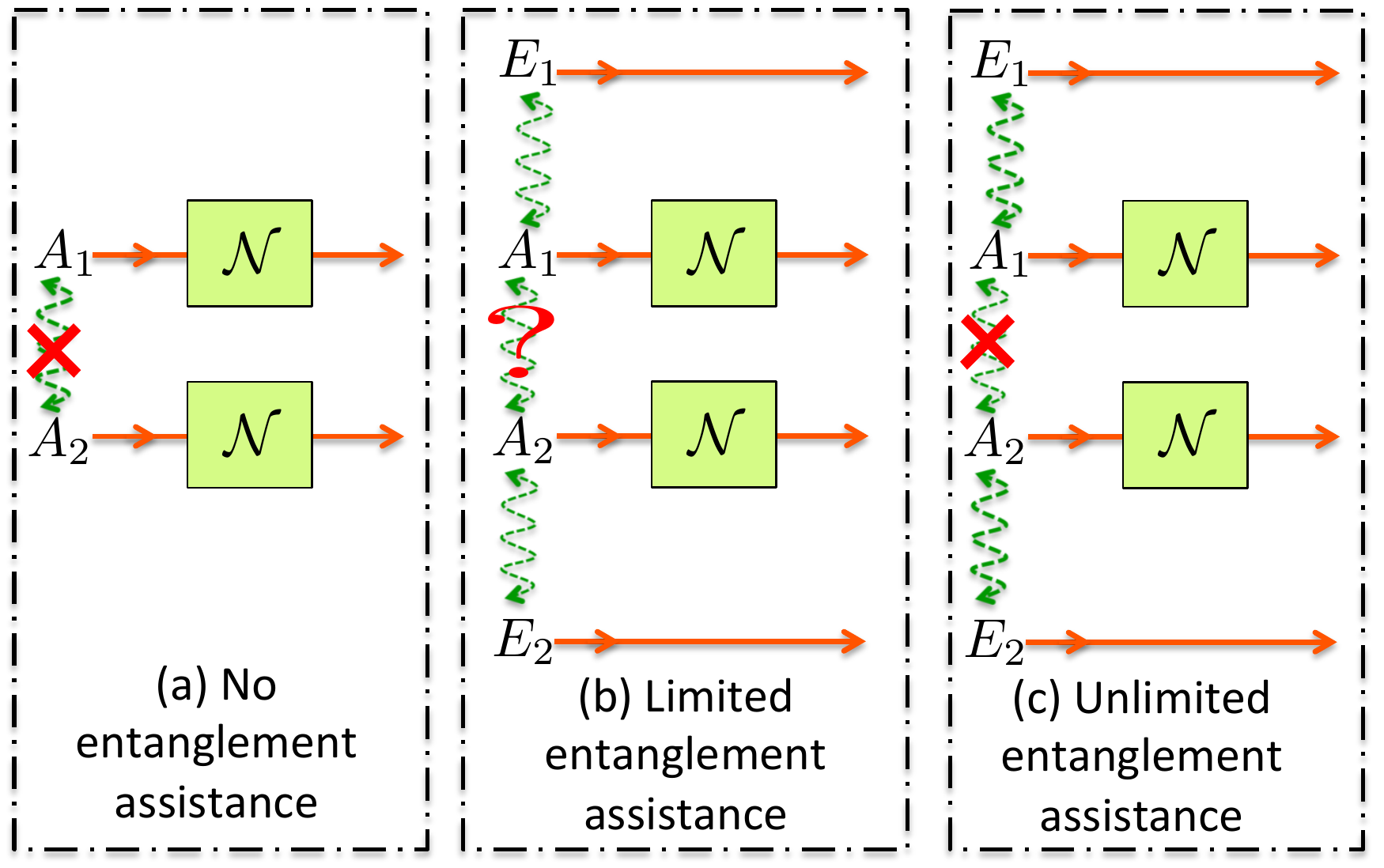}
\caption{Consider a channel $\mathcal{N}$ for classical communication, with additive classical capacity. We have the following three scenarios. (a) Entanglement across channel uses does not help if we do not have any assistance. (c) Entanglement across channel uses also does not help if we have unlimited entanglement assistance (this is always true regardless of the channel). The question addressed is case (b), whether entanglement across channel uses can help if we have some entanglement assistance.}
\label{fig:scenario}
\end{figure}
	
One can also ask if it is possible to characterize the additivity of a capacity region (\textit{e.g.,} CE trade-off) from some of its subregions (\textit{e.g.,} $ \mathcal{C} $). This has been shown to be possible for QE trade-off, as additivity of $ \mathcal{Q} $ implies the additivity of quantum capacity with any amount of entanglement assistance~\cite{Devetak08}. However, the same problem is open for CE trade-off. This question has only been recently explored~\cite{Zhuang16}, where one can restrict the encoding and constraint on entanglement to make it additive.

\begin{figure}[h]
\includegraphics[width=0.45\textwidth]
{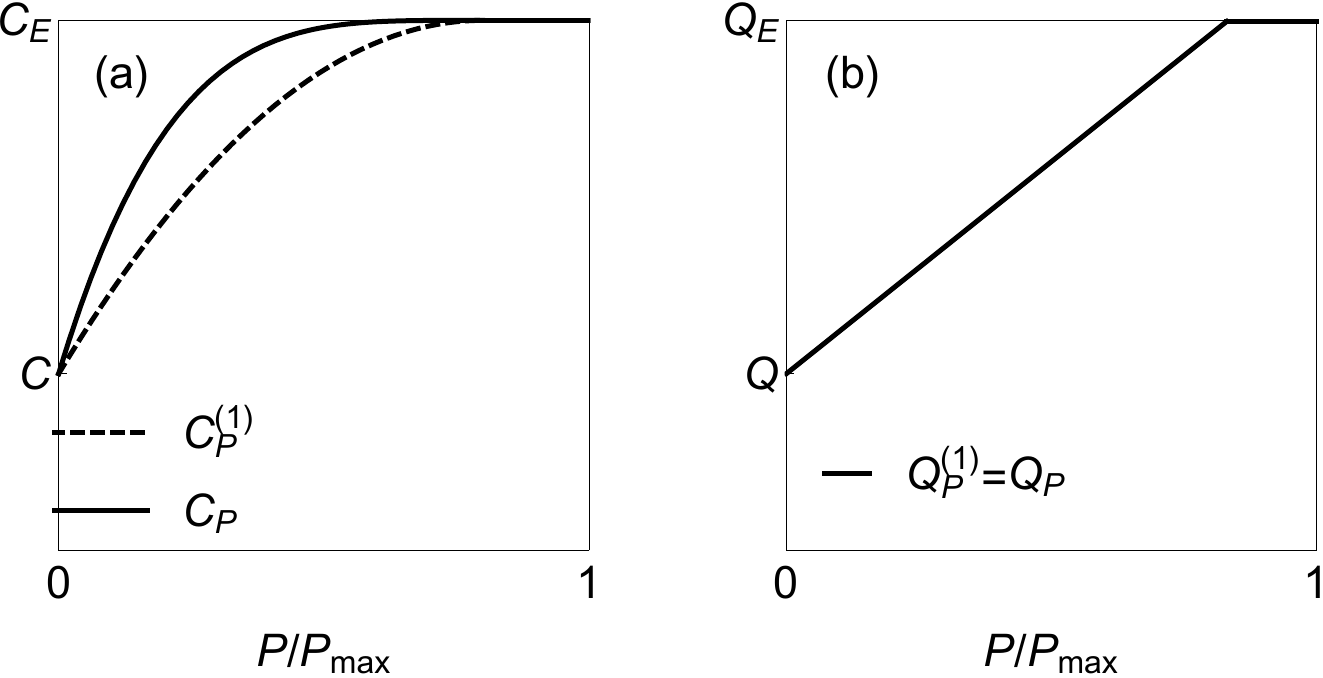}
\caption{(a) Schematic plot of the superadditivity in CE trade-off for our channel. $ \mathcal{C}_P=\mathcal{C}^{(1)}_P $ at $ P=0 $ and $ P_{\rm max} $, but not for all values in between. (b) QE trade-off curve for channels with additive quantum capacity. $ P_{\rm max} $ is the maximum amount of available entanglement assisntance.}
\label{fig:nonadd}
\end{figure}

In this work, we consider the implication of additivity of the classical capacity on the CE trade-off region.  Suppose $ \mathcal{C} $ is additive, this means we can look at each channel separately, and entangled input states do not help [Fig.~\ref{fig:scenario}(a)]. The same is true if there is unlimited entanglement assistance [Fig.~\ref{fig:scenario}(c)]. But with limited entanglement assistance, it is unclear whether entangled input states could help [Fig.~\ref{fig:scenario}(b)]. We answer the above question affirmatively. We show that there exists a channel $ \mathcal{N} $ such that the classical capacity is additive, but with some entanglement assistance $ P $, it becomes superadditive. We give a schematic plot of our CE trade-off curve in Fig.~\ref{fig:nonadd}(a).
	
To describe our results precisely, we need to first review a few key notions and results in classical capacity. To transmit classical information, Alice picks a set of signal states $ \rho_i $ with probability $ p_i $ (denoted as $\left\{p_i,\rho_i\right\}$), and sends them through the channel $ \Phi $ to Bob. The $ 1 $-shot classical capacity (\textit{i.e.,} Holevo capacity)~\cite{Holevo98,Schumacher97} of $ \Phi $ is
\be
\mathcal{C}^{(1)}\left(\Phi\right)=\max_{\{p_i,\rho_i\}}S\left(\sum_ip_i\Phi\left(\rho_i\right)\right)-\sum_ip_iS\left(\Phi\left(\rho_i\right)\right)\nonumber,
\ee 
where $ S\left(\rho\right)=-\text{tr}[\rho\log(\rho)] $ is the von Neumann entropy. This is the maximal rate of reliable classical information transmission achieved using tensor products of states $ \rho_i $, hence the ``1-shot'' classical capacity \cite{Note2}. If we can use input states which are entangled across $ n $ channel uses, we obtain  the $ n $-shot classical capacity $ \mathcal{C}^{(n)}\left(\Phi\right)=\mathcal{C}^{(1)}\left(\Phi^{\otimes n}\right)/n $. $ \mathcal{C}\left(\Phi\right)=\lim_{n\to\infty}\mathcal{C}^{(n)}\left(\Phi\right) $ denotes the (regularized) classical capacity and is the ultimate limit of reliable classical information transmission through $ \Phi $. If $ \mathcal{C}\left(\Phi\right) $ is additive for channel $ \Phi $, \textit{i.e.,} $ \mathcal{C}^{(n)}\left(\Phi\right)=\mathcal{C}^{(1)}\left(\Phi\right) $ for all $ n $, then we use $ \mathcal{C}\left(\Phi\right) $ in place of $ \mathcal{C}^{(n)}\left(\Phi\right) $.
	
Now consider the scenario where the purifications of the states $ \rho_i $ are preshared to Bob, who can use them together with the states he receives through $ \Phi $ for decoding. If we restrict the average amount of preshared entanglement to be $ P $ ebits per channel use, we arrive at the 1-shot classical capacity with entanglement assistance $ P $~\cite{Shor04}, denoted as $ \mathcal{C}^{(1)}_P\left(\Phi\right) $,
\begin{align*}
&\mathcal{C}^{(1)}_P\left(\Phi\right)=\max_{\substack{\{p_i,\rho_i\}\\ \sum_ip_iS\left(\rho_i\right)\leq P}}
\sum_ip_iS\left(\rho_i\right)\\
&+S\left(\Phi\left(\sum_ip_i\rho_i\right)\right)-\sum_ip_iS\left(\Phi\otimes\mathcal{I}\left(\phi_i\right)\right),
\end{align*}
where $ \phi_i:=\ket{\phi_i}\bra{\phi_i} $ is the density matrix of $ \rho_i $ together with a purification. This is also achieved using inputs which are tensor products of states $ \rho_i $. Similar to classical capacity, there is $ \mathcal{C}^{(n)}_P\left(\Phi\right)=\mathcal{C}^{(1)}_{nP}\left(\Phi^{\otimes n}\right)/n $ and $ \mathcal{C}_P\left(\Phi\right) $. Note that the above formula works for any $ P $. In particular, when $ P=0 $, we get $ \mathcal{C}^{(1)}\left(\Phi\right) $. When $ P$ is maximal, we get $ \mathcal{C}_E\left(\Phi\right) $.
	
Now we are ready to state our main result.
\begin{theorem}
(Main Theorem) There exists a channel $ \mathcal{N} $ such that
\be
\mathcal{C}\left(\mathcal{N}\right)=\mathcal{C}^{(1)}\left(\mathcal{N}\right)\nonumber
\ee
\textit{i.e.,} its classical capacity is additive. However, there exists $ P $ such that
\be
\mathcal{C}_P\left(\mathcal{N}\right)>\mathcal{C}^{(1)}_P\left(\mathcal{N}\right)\nonumber
\ee
\textit{i.e.,} its classical capacity with limited entanglement assistance can be superadditive.
\end{theorem}
This additivity to superadditivity transition in classical capacity is illustrated in Fig.~\ref{fig:nonadd}(a). This is in sharp contrast to the QE trade-off curve [Fig.~\ref{fig:nonadd}(b)], as $ \mathcal{Q}_P^{(n)} $ grows linearly in $ P $ with gradient 1. Additivity of $ \mathcal{Q}_p $ follows from the additivity of $ \mathcal{Q} $.

\begin{figure}
\includegraphics[width=0.4\textwidth]
{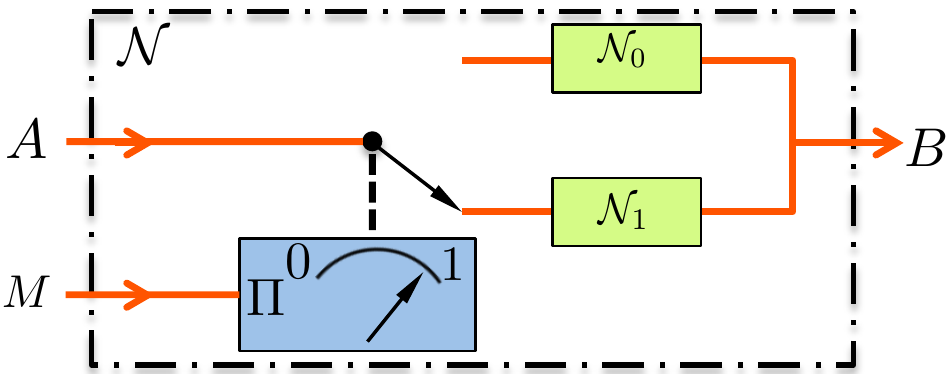}
\caption{Diagrammatic representation of $ \mathcal{N} $.}
\label{fig:schematicN}
\end{figure}

Our channel $ \mathcal{N} $ is a conditional quantum channel $ \mathcal{N}^{MA\to B} $~\cite{Yard05}, where register $ M $ determines whether $ \mathcal{N}_0^{A\to B} $ or $ \mathcal{N}_1^{A\to B} $ is used (see Fig.~\ref{fig:schematicN} for a diagrammatic representation). Explicitly, on any input state $ \rho^{MA} $ \cite{Note3},
\be\label{eq:N}
\mathcal{N}(\rho^{MA})
=\mathcal{N}_0\left(\bra{0}\rho^{MA}\ket{0}^M\right)+ \mathcal{N}_1\left(\bra{1}\rho^{MA}\ket{1}^M\right).
\ee
This construction is similar to the one in Ref. \cite{Elkouss15}. However, their construction does not directly apply to our case since $M$ is kept and contains classical information.
	
The intuition why a channel like $ \mathcal{N} $ will work is that without entanglement, we are only using the classical channel $ \mathcal{N}_0 $, hence its classical capacity is additive. As one increases entanglement assistance, one starts using the quantum channel $ \mathcal{N}_1 $, where superadditivity kicks in.
	
Our construction is generic and does not depend on the specific forms of $ \mathcal{N}_0 $ and $ \mathcal{N}_1 $. Hence we give the properties of $ \mathcal{N}_0 $ and $ \mathcal{N}_1 $ that are required for our argument to work and will give a construction of $ \mathcal{N}_1 $ later. An example of $ \mathcal{N}_0 $ is given in the Supplemental Material.
	
We require the classical channel $ \mathcal{N}_0 $ to have the following properties: (0.1) $ \mathcal{C}\left(\mathcal{N}_0\right)=\log(|B|)-\min S\left(\mathcal{N}_0\left(\rho\right)\right) $.
(0.2) It has a noise parameter $ \eta $ which can be tuned, such that $ \mathcal{C}\left(\mathcal{N}_0\right) $ varies from $ 0 $ to $ \log(|B|) $ continuously.

We require the quantum channel $ \mathcal{N}_1 $ to have the following properties: (1.1) It has a superadditive classical capacity, \textit{i.e.,}	$ \mathcal{C}\left(\mathcal{N}_1\right)>\mathcal{C}^{(1)}\left(\mathcal{N}_1\right) $.
(1.2) For any $ n $ and $ P $,
\be
\mathcal{C}^{(n)}_P(\mathcal{N}_1)=\log (|B|)-\min_{S(\rho)\leq nP} \frac{1}{n}\{S\left(\mathcal{N}_1^{\otimes n}\otimes \mathcal{I}\left(\phi_\rho\right)\right)-S\left(\rho\right)\}.\nonumber
\ee
(1.3) There exists $ P>0 $ such that $ \mathcal{C}_P\left(\mathcal{N}_1\right)>\mathcal{C}^{(1)}_{P}\left(\mathcal{N}_1\right) $, and $ \mathcal{C}_P\left(\mathcal{N}_1\right) $ is strictly concave at $ P $.
	
Here by saying a function $ f $ is strictly concave at $ y $, we mean $ f(y)>(1-p)f(v)+pf(w) $ for all $ v<y<w $ satisfying $ (1-p)v+pw=y $, with $ p\in(0,1) $. It is clear that $ \mathcal{C}_P(\Phi) $ is always concave in $ P $. If $ P=(1-p)P_1+pP_2 $, then $ \mathcal{C}_P(\Phi)\geq p\mathcal{C}_{P_1}(\Phi)+(1-p)\mathcal{C}_{P_2}(\Phi) $, as one can always just use entanglement $ P_1 $ for the $ p $ fraction of the channel uses and entanglement $ P_2 $ for the other fraction.
	
The rest of the paper is organized as follows. We first state Lemma~\ref{lemma:cq} about the classical capacity with limited entanglement assistance, of partial cq channels (defined in Lemma~\ref{lemma:cq}). This lemma together with the properties above lead to the simplification of capacity formulas, as we show in Lemmas~\ref{lemma:C} and \ref{lemma:CP}. We will prove our main theorem in the main text and leave the proofs of various lemmas to the Supplemental Material.
	
\begin{lemma}\label{lemma:cq}
Suppose a channel $ \Psi $ has an input Hilbert space $ \mathcal{H}_R\otimes\mathcal{H}_C $. If there exists a noiseless classical channel $ \Pi $ on $ \mathcal{H}_R $ with orthonormal basis $ \{\ket{j}\} $, such that
\be
\Psi=\Psi\circ \left(\Pi\otimes\mathcal{I}_C\right)\nonumber,
\ee
then $ \mathcal{C}^{(1)}_P(\Psi) $ can be achieved with an input ensemble $ \{p_{ij},\ket{j}\bra{j}\otimes\rho_{ij}\} $, where $ \rho_{ij} $ are states of $ C $.
\end{lemma}
By saying $ \Pi $ is a noiseless classical channel with orthonormal basis $ \{\ket{j}\} $, we mean $ \Pi\left(\rho\right)=\sum_j\ket{j}\bra{j}\rho\ket{j}\bra{j} $. This is very intuitive. Entanglement between $ R $ and other parties is not useful, as $ \Pi $ destroys it. Since we only have limited entanglement, it is better to use it on $ C $.
	
Using Lemma~\ref{lemma:cq} and properties of $ \mathcal{N}_0 $ and $ \mathcal{N}_1 $, we can simplify the various capacity formulas of $ \mathcal{N} $.
\begin{lemma}\label{lemma:C}
\begin{align*}
\mathcal{C}^{(1)}(\mathcal{N})&= \max\left\{\mathcal{C}\left(\mathcal{N}_0\right),\mathcal{C}^{(1)}\left(\mathcal{N}_1\right)\right\},\\
\mathcal{C}(\mathcal{N})&= \max\left\{\mathcal{C}(\mathcal{N}_0),\mathcal{C}(\mathcal{N}_1)\right\}.
\end{align*}    
\end{lemma}
Lemma~\ref{lemma:cq} ensures that for different uses of the channel, we can choose to use $ \mathcal{N}_0 $ or $ \mathcal{N}_1 $ only, without sacrificing the capacity. Lemma~\ref{lemma:C} simply states that, for all channel uses, we should use either $ \mathcal{N}_0 $ or $ \mathcal{N}_1 $ .
\begin{lemma}\label{lemma:CP}
\begin{align}
\mathcal{C}^{(1)}_P(\mathcal{N})&=\max_{\substack{{\{q,P'\}}\\{(1-q)P'=P}}}q\mathcal{C}(\mathcal{N}_0)+(1-q)\mathcal{C}^{(1)}_{P'}(\mathcal{N}_1)\label{eq:CP1shotgen},\\
\mathcal{C}_P\left(\mathcal{N}\right)&=\max_{\substack{{\{q,P'\}}\\{(1-q)P'=P}}}q\mathcal{C}\left(\mathcal{N}_0\right)+(1-q)\mathcal{C}_{P'}\left(\mathcal{N}_1\right).\label{eq:CPgen}
\end{align}
\end{lemma}
This lemma states that,  for entanglement-assisted classical communication,  the best strategy is to use $ \mathcal{N}_0 $ for some fraction of the channel uses and $ \mathcal{N}_1 $ for the other fractions of the channel uses (\textit{i.e.,} time sharing). Since using $ \mathcal{N}_0 $ does not require entanglement assistance, we can allocate more of it to $ \mathcal{N}_1 $.
	
Now we are ready to prove the main theorem.
{\bf Proof of main theorem---} Choose $ \mathcal{N}_0 $ such that
\be\label{eq:N0capacity}
\mathcal{C}\left(\mathcal{N}_0\right)=\mathcal{C}\left(\mathcal{N}_1\right)>\mathcal{C}^{(1)}\left(\mathcal{N}_1\right).
\ee
By Lemma~\ref{lemma:C}, the classical capacity of $ \mathcal{N} $ is additive, \textit{i.e.,}
\be
\mathcal{C}\left(\mathcal{N}\right)=\mathcal{C}\left(\mathcal{N}_0\right)=\mathcal{C}^{(1)}\left(\mathcal{N}\right).
\ee
From Eqs.~(\ref{eq:CPgen}),(\ref{eq:N0capacity}) and concavity of $ \mathcal{C}_P\left(\mathcal{N}_1\right) $ with respect to $ P $, we have $
\mathcal{C}_P\left(\mathcal{N}\right)\leq\mathcal{C}_P\left(\mathcal{N}_1\right) $. Also, $ \mathcal{C}_P\left(\mathcal{N}\right)\ge\mathcal{C}_P\left(\mathcal{N}_1\right) $ by choosing $ q=0 $ in Eq.~(\ref{eq:CPgen}).
So we have
\be\label{eq:CPN}
\mathcal{C}_P\left(\mathcal{N}\right)=\mathcal{C}_P\left(\mathcal{N}_1\right).
\ee
Choose $ P>0 $ according to property 1.3. By Lemma~\ref{lemma:CP}, suppose $ \mathcal{C}^{(1)}_P\left(\mathcal{N}\right) $ is achieved at some $ \left\{\tilde{q},\tilde{P}\right\} $ with $ \left(1-\tilde{q}\right)\tilde{P}=P $, \textit{i.e.,}
\be
\mathcal{C}^{(1)}_P(\mathcal{N})=\tilde{q}\mathcal{C}\left(\mathcal{N}_0\right)+\left(1-\tilde{q}\right)\mathcal{C}^{(1)}_{\tilde{P}}\left(\mathcal{N}_1\right).
\ee
If $ \tilde{P}=P $, we have
\be
\mathcal{C}_P\left(\mathcal{N}\right)=\mathcal{C}_P\left(\mathcal{N}_1\right)>\mathcal{C}^{(1)}_{P}\left(\mathcal{N}_1\right)=\mathcal{C}^{(1)}_{P}\left(\mathcal{N}\right),
\ee
where the inequality follows from property 1.3.\\
If $ \tilde{P}>P $ and thus $ \tilde{q}>0 $,
\begin{align}
\mathcal{C}_P\left(\mathcal{N}\right)&=\mathcal{C}_P\left(\mathcal{N}_1\right)>\tilde{q}\mathcal{C}\left(\mathcal{N}_1\right)+\left(1-\tilde{q}\right)\mathcal{C}_{\tilde{P}}\left(\mathcal{N}_1\right)\nonumber\\
&\geq \tilde{q}\mathcal{C}\left(\mathcal{N}_0\right)+\left(1-\tilde{q}\right)\mathcal{C}^{(1)}_{\tilde{P}}\left(\mathcal{N}_1\right)=\mathcal{C}^{(1)}_P\left(\mathcal{N}\right),
\end{align}
where the first inequality follows from property 1.3.
	
{\bf Construction of $ \mathcal{N}_1 $---}The first two properties for $ \mathcal{N}_1 $ can be easily satisfied. One can take a channel with a subadditive minimum output entropy \cite{Hastings09} and unitally extend it to a channel with a superadditive classical capacity, via Shor's construction~\cite{Shor04additivity,Fukuda07}. Unfortunately, such channels are poorly understood, and we do not know if it satisfies property 1.3. We argue that if it does not, we can tensor product a dephasing channel that will guarantee it is satisfied, without sacrificing the other properties.
	
We quote the following property about concave functions \cite{Rudin76}:
A concave function $ u(y) $ is continuous, differentiable from the left and from the right. The derivative is decreasing, \textit{i.e.,} for $ x<y $, we have $ u'(x-)\geq u'(x+)\geq u'(y-)\geq u'(y+) $.
We use ``$ \pm $'' to denote the right and left derivatives when needed.
	
Let $ \mathcal{E}^{C\to C} $ be a random orthogonal channel with subadditive minimum output entropy~\cite{Hastings09} and $ \mathcal{F}^{RC\to C} $ (with $ |R|=|C|^2 $) be a conditional quantum channel of the form
\be
\mathcal{F}\left(\rho^{RC}\right)=\sum_{j=1}^{|C|^2} X_j\mathcal{E}\left(\bra{j}\rho^{RC}\ket{j}^R\right)X_j^\dagger,
\ee
where $ X_j $'s are the Heisenberg-Weyl operators on $ C $ \cite{Shor02}.
This ensures $ \mathcal{F} $ satisfies properties 1.1 and 1.2~\cite{Note4}.
	
Because of Lemma~\ref{lemma:cq}, the useful entanglement assistance is at most $ \log(|C|) $. Thus, we restrict to $ 0\leq P\leq \log(|C|) $.
	
Let
\be
\epsilon=\mathcal{C}\left(\mathcal{F}\right)-\mathcal{C}^{(1)}\left(\mathcal{F}\right)>0\label{eq:gap}.
\ee
Since 
\be
\mathcal{C}^{(1)}_P\left(\mathcal{F}\right)\leq \mathcal{C}^{(1)}\left(\mathcal{F}\right)+P\label{eq:CPub},
\ee
\be
\mathcal{C}_E\left(\mathcal{F}\right)\leq \mathcal{C}\left(\mathcal{F}\right)+\log(|C|)-\epsilon.
\ee
This implies $ d\mathcal{C}_P\left(\mathcal{F}\right)/dP $ cannot always be 1. Thus, there exists $ \bar{P}\in[0,\log(|C|)) $ such that
\be
d\mathcal{C}_P\left(\mathcal{F}\right)/dP=1,~~\forall~ 0\leq P\leq\bar{P}
\ee
and
\be
d\mathcal{C}_P\left(\mathcal{F}\right)/dP<1,~~\forall P>\bar{P}.
\ee
Next, we discuss the few different cases.
(1) $ \bar{P}>0 $. Then $ \mathcal{C}_P\left(\mathcal{F}\right) $ is strictly concave at $ \bar{P} $ by definition.	Note that $ \mathcal{C}_{\bar{P}}\left(\mathcal{F}\right)=\mathcal{C}\left(\mathcal{F}\right)+\bar{P} $
but $ \mathcal{C}^{(1)}_{\bar{P}}\left(\mathcal{F}\right)\leq \mathcal{C}^{(1)}\left(\mathcal{F}\right)+\bar{P} $,
thus $ \mathcal{C}_{\bar{P}}\left(\mathcal{F}\right)-\mathcal{C}^{(1)}_{\bar{P}}\left(\mathcal{F}\right)\geq \epsilon $ and $ \mathcal{N}_1=\mathcal{F} $ satisfies property 1.3.
(2)	$ \bar{P}=0 $. Let $ \mathcal{N}_1=\mathcal{F}\otimes\Delta^Z_\lambda $, where $ \Delta^Z_\lambda $ is the qubit dephasing channel $ \Delta^Z_\lambda\left(\rho\right)=(1-\lambda)\rho+\lambda Z\rho Z $. The CQE trade-off region is additive for $ \Phi\otimes\Delta^Z_\lambda $, for any channel $ \Phi $; thus, $ \mathcal{N}_1 $ satisfies property 1.1. $ \Delta^Z_\lambda $ satisfies property 1.2, and by arguments similar to Appendix B of Ref.~\cite{Bradler10}, one can show $ \mathcal{N}_1 $ also satisfies property 1.2.
	
Since $	d\mathcal{C}_P\left(\mathcal{F}\right)/dP|_{0+}$$<1$,
choose $ \lambda>0 $ small such that $
d\mathcal{C}_P\left(\Delta^Z_\lambda\right)/dP|_{1-}$$>d\mathcal{C}_P\left(\mathcal{F}\right)/dP|_{0+} $.
This ensures that when $ 0<P\leq 1$,
\be
\mathcal{C}_P\left(\mathcal{N}_1\right)=\mathcal{C}\left(\mathcal{F}\right)+\mathcal{C}_P\left(\Delta^Z_\lambda\right)\label{eq:CPadd}.
\ee
Since $ \mathcal{C}_P\left(\Delta^Z_\lambda\right) $ is strictly concave with respect to $ P $ when $ \lambda<1/2 $ \cite{Hsieh10}, $ \mathcal{C}_P\left(\mathcal{N}_1\right) $ is also strictly concave with respect to $ P $, for $ 0<P\leq 1 $. Also, when $ P<\epsilon $,
\begin{align*}
\mathcal{C}_P\left(\mathcal{N}_1\right)&\geq \mathcal{C}\left(\mathcal{F}\right)+\mathcal{C}\left(\Delta^Z_\lambda\right)\\
&>\mathcal{C}^{(1)}\left(\mathcal{F}\right)+\mathcal{C}\left(\Delta^Z_\lambda\right)+P\geq\mathcal{C}^{(1)}_P\left(\mathcal{N}_1\right),
\end{align*}
where the first inequality comes from Eq.(\ref{eq:CPadd}), the second one comes from our assumption $ P<\epsilon $ and Eq. (\ref{eq:gap}) and the last one comes from Eq.(\ref{eq:CPub}).\\		
This ensures that $ \mathcal{C}_P\left(\mathcal{N}_1\right) $ is superadditive. Thus when $ 0<P<\min\{1,\epsilon\}$, $ \mathcal{C}_P\left(\mathcal{N}_1\right) $ is strictly concave and superadditive, satisfying property 1.3.\\
	
{\em Conclusion.---}Our work unveils the complications in characterizing the additivity of the CE capacity region. In fact, the only known channels that admit an additive CE capacity region are the quantum erasure channels \cite{Hsieh10} and Hadamard channels \cite{Bradler10}, many fewer than the class of channels with an additive classical capacity. Coincidentally, these two classes of channels also admit an additive CQE trade-off capacity, suggesting a nontrivial connection \cite{Hsieh10,Bradler10,Zhu17}.  
	
Also, we do not know the number of shots at which the superadditivity occurs. However, it is very likely that our $ \mathcal{N}_1 $ only has superadditivity in classical capacity up to two shots\cite{Montanaro13}. In that case, the superadditivity in classical capacity with limited entanglement will appear at two shots.
	
EYZ and QZ would like to thank Min-Hsiu Hsieh for many insightful discussions. EYZ and PWS are supported by the National Science Foundation under Grant Contract No. CCF-1525130. QZ is supported by the Claude E. Shannon Research Assistantship.  PWS is supported by the NSF through the STC for Science of Information under Grant No. CCF0-939370.
%
%

\pagebreak

\setcounter{equation}{0}
\setcounter{figure}{0}
\setcounter{table}{0}
\makeatletter
\renewcommand{\theequation}{S\arabic{equation}}
\renewcommand{\thefigure}{S\arabic{figure}}
\renewcommand{\bibnumfmt}[1]{[S#1]}
\renewcommand{\citenumfont}[1]{S#1}
\widetext
\begin{center}
\large\textbf{Supplemental Materials: Superadditivity of the Classical Capacity with Limited Entanglement Assistance}

\end{center}
\subsection{Proof of Lemma~\ref{lemma:cq} in Main Text}
\begin{proof}
For channel $ \Psi $, suppose one uses the ensemble $ \left\{p_i,\rho_i \right\} $ for entanglement-assisted classical communication. Holevo's bound gives the maximum classical information transmitted as
\be\label{eq:pre}
\chi_{\rm assist}\left(\Psi,\left\{p_i,\rho_i \right\}\right):=\sum_ip_iS\left(\rho_i\right)+S\left(\Psi\left(\sum_ip_i\rho_i\right)\right)-\sum_ip_iS\left(\Psi\otimes\mathcal{I}_E\left(\phi_i\right)\right),
\ee
where the pre-shared entanglement is in Hilbert space $ \mathcal{H}_E $, and $ \ket{\phi_i}\in \mathcal{H}_R\otimes\mathcal{H}_C\otimes\mathcal{H}_E $. Then
\be
\mathcal{C}_P^{(1)}\left(\Psi\right)=\max_{\substack{\{p_i,\rho_i\}\\ \sum_ip_iS\left(\rho_i\right)\leq P}}\chi_{\rm assist}\left(\Psi,\left\{p_i,\rho_i \right\}\right).
\ee	
We apply the noiseless classical channel $ \Pi $ on the register $ R $. Since $ \Psi=\Psi\circ \left(\Pi\otimes\mathcal{I}_C\right) $, this does not change the amount of information transmitted.
Hence Eq.(\ref{eq:pre}) is equal to
\be\label{eq:pm}
\sum_ip_iS\left(\rho_i\right)+S\left(\Psi\left(\sum_ip_i\Pi\otimes\mathcal{I}_C\left(\rho_i\right)\right)\right)-\sum_ip_iS\left(\Psi\otimes\mathcal{I}_E\left(\Pi\otimes\mathcal{I}_{CE}\left(\phi_i\right)\right)\right).
\ee

Now, we consider an alternative protocol, described below. Formally, the state of $RCE$ after $ \Pi $ is
\be\label{eq:postPistate}
\Pi\otimes\mathcal{I}_{CE}\left(\phi_i\right)=\sum_jp(j|i)\ket{j}\bra{j}\otimes\rho_{ij}.
\ee
Note $\rho_{ij}$ are states of CE and $\ket{j}$'s are the basis of the classical channel. Moreover, for the density matrix $ \rho_{ij} $, consider its spectral decomposition
\be
\rho_{ij}=\sum_k p(k|ij)\phi_{ijk}.
\ee
So Eq.~(\ref{eq:postPistate}) is equal to
\begin{align}
\Pi\otimes\mathcal{I}_{CE}\left(\phi_i\right)&=\sum_jp(j|i)\ket{j}\bra{j}\otimes\left(\sum_k p(k|ij)\phi_{ijk}\right)\nonumber\\
&=\sum_{jk}p(jk|i)\ket{j}\bra{j}\otimes\phi_{ijk},
\end{align}
where $ p(jk|i)=p(k|ij)p(j|i) $. We introduce the notation for the reduced density matrices $ \rho^C_{ijk}=\mathrm{tr}_E\left(\phi_{ijk}\right) $ and $ \rho^E_{ijk}=\mathrm{tr}_C\left(\phi_{ijk}\right) $. Denote $ p_{ijk}=p(jk|i)p_i $.

Suppose Alice and Bob instead use the ensemble $ \left\{p_{ijk},\ket{j}\bra{j}\otimes\rho^C_{ijk}\right\} $ for entanglement-assisted classical communication. We prove in the following that using this ensemble will consume less entanglement and transmit more information, which suffices to prove the lemma.

First, we show that the entanglement consumption for the second protocol is less than that for the original protocol. The original entanglement assistance for using state $\phi_i$ is
\be
\rho_i^E=\mathrm{tr}_{CR}\left(\phi_i\right)=\mathrm{tr}_{CR}\left(\Pi\otimes\mathcal{I}_{CE}\left(\phi_i\right)\right)=\sum_jp(jk|i)\rho^E_{ijk},
\ee
where $\rho^E_{ijk}$'s are the entanglement assistance for the second protocol.
By concavity of the von Neumann entropy,
\be
\sum_{jk}p(jk|i)S\left(\rho^E_{ijk}\right)\leq S\left(\rho_i\right),
\ee
and thus
\be
\sum_{ijk}p_{ijk}S\left(\rho^E_{ijk}\right)\leq\sum_ip_iS\left(\rho_i\right).
\ee
This means the average entanglement used between Alice and Bob is smaller for the second protocol.

Next, we show that the amount of information transmitted by the second protocol is larger than that by the first protocol.
For the second protocol, $ \chi_{\rm assist}\left(\Psi,\left\{p_{ijk},\ket{j}\bra{j}\otimes\rho^C_{ijk}\right\}\right) $ is
\be\label{eq:post}
\sum_{ijk}p_{ijk}S\left(\rho^E_{ijk}\right)+S\left(\Psi\left(\sum_{ijk}p_{ijk}\ket{j}\bra{j}\otimes\rho^C_{ijk}\right)\right)-\sum_{ijk}p_{ijk}S\left(\Psi\otimes\mathcal{I}_E\left(\ket{j}\bra{j}\otimes\phi_{ijk}\right)\right).
\ee
We introduce a new register $Q$, which records the $j,k$ indices of the second ensemble of states. Denote each state as $ \ket{jk}\in \mathcal{H}_Q $. Denote $C'$ the output register of the channel $\Psi$. Consider the following state of $QC'E$
\be
\sum_{jk}p(jk|i)\ket{jk}\bra{jk}\otimes\left(\Psi\otimes\mathcal{I}_E\left(\ket{j}\bra{j}\otimes\phi_{ijk}\right)\right).
\ee
It has the following properties,
\begin{subequations}
\begin{align}
&S\left(E\right)=S\left(\sum_{jk}p(jk|i)\rho^E_{ijk}\right)=S\left(\rho_i\right),\\
&S\left(C'E\right)=S\left(\sum_{jk}p(jk|i)\Psi\otimes\mathcal{I}_E\left(\ket{j}\bra{j}\otimes\phi_{ijk}\right)\right)=S\left(\Psi\otimes\mathcal{I}_E\left(\Pi\otimes\mathcal{I}_{CE}\left(\phi_i\right)\right)\right),\\
&S\left(QE\right)-S\left(Q\right)=S\left(\sum_{jk}p(jk|i)\ket{jk}\bra{jk}\otimes\rho^E_{ijk}\right)-S\left(\sum_{jk}p(jk|i)\ket{jk}\bra{jk}\right)=p_i^{-1}\sum_{jk}p_{ijk}S\left(\rho^E_{ijk}\right),\\
&S\left(QC'E\right)-S\left(Q\right)=S\left(\sum_{jk}p(jk|i)\ket{jk}\bra{jk}\otimes\left(\Psi\otimes\mathcal{I}_E\left(\ket{j}\bra{j}\otimes\phi_{ijk}\right)\right)\right)-S\left(\sum_{jk}p(jk|i)\ket{jk}\bra{jk}\right)\nonumber\\
&~~~~~~~~~~~~~~~~~~~~~~~=p_i^{-1}\sum_{jk}p_{ijk}S\left(\Psi\otimes\mathcal{I}_E\left(\ket{j}\bra{j}\otimes\phi_{ijk}\right)\right).
\end{align}
\label{eq:interpretation}
\end{subequations}
By strong subadditivity,
\be
S\left(QE\right)-S\left(Q\right)-\left(S\left(QC'E\right)-S\left(Q\right)\right)\geq S\left(E\right)-S\left(C'E\right).
\label{eq:ssa}
\ee

Now we are ready to compare Eqs.~(\ref{eq:pm}) and (\ref{eq:post}). The second term of Eq.~(\ref{eq:pm}) and Eq.~(\ref{eq:post}) are the same. From Eq.~(\ref{eq:ssa}), with each term given in Eqs.~(\ref{eq:interpretation}), we immediately see that the first and the third terms of Eq.~(\ref{eq:pm}) and Eq.~(\ref{eq:post}) are exactly the terms in Eq.~(\ref{eq:ssa}).
So Eq.~(\ref{eq:post}) is greater than Eq.~(\ref{eq:pm}).
\end{proof}

\subsection{proof of Lemma~\ref{lemma:C} in Main Text}
\begin{proof}
Let $ \Pi(\rho)=\ket{0}\bra{0}\rho\ket{0}\bra{0}+\ket{1}\bra{1}\rho\ket{1}\bra{1} $. Then $ \mathcal{N}=\mathcal{N}\circ\left(\Pi\otimes\mathcal{I}_A\right) $ and $ \mathcal{N} $ is a partial cq channel.
Taking $ P=0 $ in Lemma~\ref{lemma:cq} from the main text, $ \mathcal{C}^{(1)}(N) $ can be achieved with ensembles of the form $ \left\{p_{ij},\ket{j}\otimes\ket{\phi_{ij}} \right\} $, with $ j\in \{0,1\} $.

The Holevo information of $ \mathcal{N} $ with respect to $ \left\{p_{ij},\ket{j}\otimes\ket{\phi_{ij}} \right\} $ is given by
\begin{align}
&\chi\left(\mathcal{N},\left\{p_{ij},\ket{j}\otimes\ket{\phi_{ij}} \right\}\right)\nonumber\\
=&S\left(\sum_i p_{i0}\mathcal{N}_0\left(\phi_{i0}\right)+p_{i1}\mathcal{N}_1\left(\phi_{i1}\right)\right)-\sum_i p_{i0}S\left(\mathcal{N}_0\left(\phi_{i0}\right)\right)-\sum_ip_{i1}S\left(\mathcal{N}_1\left(\phi_{i1}\right)\right)\nonumber\\
\le& \log(|B|)-\sum_i p_{i0}S\left(\mathcal{N}_0\left(\phi_{i0}\right)\right)-\sum_ip_{i1}S\left(\mathcal{N}_1\left(\phi_{i1}\right)\right)\nonumber\\
\leq& P_0\left(\log(|B|)-\sum_i\frac{p_{i0}}{P_0}S\left(\mathcal{N}_0\left(\phi_{i0}\right)\right)\right)+P_1\left(\log(|B|)-\sum_i\frac{p_{i1}}{P_1}S\left(\mathcal{N}_1\left(\phi_{i1}\right)\right)\right)\nonumber\\
\leq& P_0\mathcal{C}(\mathcal{N}_0)+P_1\mathcal{C}^{(1)}(\mathcal{N}_1)
\end{align}
where we define $ P_0=\sum_ip_{i0} $ and $ P_1=\sum_ip_{i1} $. In the last line, we used properties (0.1) and (1.2) from the main text.
This means
\be
\mathcal{C}^{(1)}(\mathcal{N})\leq\max\{\mathcal{C}(\mathcal{N}_0),\mathcal{C}^{(1)}(\mathcal{N}_1)\}.
\ee
On the other hand, by choosing states such that the first register is 0 or 1, it is obvious that
\be
\mathcal{C}^{(1)}(\mathcal{N})\geq\max\{\mathcal{C}(\mathcal{N}_0),\mathcal{C}^{(1)}(\mathcal{N}_1)\},
\ee
so
\be
\mathcal{C}^{(1)}(\mathcal{N})=\max\{\mathcal{C}(\mathcal{N}_0),\mathcal{C}^{(1)}(\mathcal{N}_1)\}.
\ee
Similarly
\be
\mathcal{C}^{(n)}(\mathcal{N})= \max_{0\leq l\leq n}\left\{\frac{1}{n}\mathcal{C}^{(1)}\left(\mathcal{N}_0^{\otimes n-l}\otimes\mathcal{N}_1^{\otimes l}\right)\right\}.
\ee
Since $ \mathcal{N}_0 $ is classical~\cite{Shor02EB_Sup},
\be
\mathcal{C}^{(1)}\left(\mathcal{N}_0^{\otimes n-l}\otimes\mathcal{N}_1^{\otimes l}\right)=(n-l)\mathcal{C}(\mathcal{N}_0)+\mathcal{C}^{(1)}\left(\mathcal{N}_1^{\otimes l}\right)=(n-l)\mathcal{C}(\mathcal{N}_0)+l\mathcal{C}^{(l)}\left(\mathcal{N}_1\right).
\ee
This means
\be
\mathcal{C}^{(n)}(\mathcal{N})= \max_{0\leq l\leq n}\left\{\frac{n-l}{n}\mathcal{C}\left(\mathcal{N}_0\right)+\frac{l}{n}\mathcal{C}^{(l)}\left(\mathcal{N}_1\right)\right\}.
\ee
Therefore,
\be
\mathcal{C}(\mathcal{N})= \max\left\{\mathcal{C}(\mathcal{N}_0),\mathcal{C}(\mathcal{N}_1)\right\}.
\ee
\end{proof}

\subsection{proof of Lemma~\ref{lemma:CP} in Main Text}
\begin{proof}
One can apply Lemma~\ref{lemma:cq} from the main text to $ \mathcal{N} $ and only consider ensembles of the form $ \left\{p_{ij},\ket{j}\bra{j}\otimes\rho_{ij}\right\} $, with $ \sum_{ij}p_{ij}S\left(\ket{j}\bra{j}\otimes\rho_{ij}\right)\leq P $. For such an ensemble, $ \chi_{\rm assist}\left(\mathcal{N},\left\{p_{ij},\ket{j}\bra{j}\otimes\rho_{ij}\right\}\right) $ is given by
\begin{align}
&\sum_i p_{i0}S\left(\ket{0}\bra{0}\otimes\rho_{i0}\right)+p_{i1}S\left(\ket{1}\bra{1}\otimes\rho_{i1}\right)+S\left(\mathcal{N}\left(\sum_i p_{i0}\ket{0}\bra{0}\otimes\rho_{i0}+\sum_i p_{i1}\ket{1}\bra{1}\otimes\rho_{i1}\right)\right)\nonumber\\
-&\sum_ip_{i0}S\left(\mathcal{N}_0\otimes\mathcal{I}\left(\phi_{i0}\right)\right)-\sum_ip_{i1}S\left(\mathcal{N}_1\otimes\mathcal{I}\left(\phi_{i1}\right)\right)\nonumber\\
\leq&P_0\left(\sum_i\frac{p_{i0}}{P_0}S\left(\rho_{i0}\right)+\log|B|-\sum_i\frac{p_{i0}}{P}S\left(\mathcal{N}_0\otimes\mathcal{I}\left(\phi_{i0}\right)\right)\right)\nonumber\\
+&P_1\left(\sum_i\frac{p_{i1}}{P_1}S\left(\rho_{i1}\right)+\log|B|-\sum_i\frac{p_{i1}}{P_1}S\left(\mathcal{N}_1\otimes\mathcal{I}\left(\phi_{i1}\right)\right)\right)\label{eq:CP1shot},
\end{align}
where $ P_0=\sum_ip_{i0} $ and $ P_1=\sum_ip_{i1} $.

Since $ \mathcal{N}_0 $ is classical, by the same argument in the proof of Lemma~\ref{lemma:cq} from the main text,
\be
\sum_i\frac{p_{i0}}{P_0}S\left(\rho_{i0}\right)+\log(|B|)-\sum_i\frac{p_{i0}}{P}S\left(\mathcal{N}_0\otimes\mathcal{I}\left(\phi_{i0}\right)\right)\leq \mathcal{C}\left(\mathcal{N}_0\right).
\ee
This means Eq.~(\ref{eq:CP1shot}) is less than
\be
P_0\mathcal{C}\left(\mathcal{N}_0\right)+P_1\mathcal{C}^{(1)}_{P/P_1}\left(\mathcal{N}_1\right),
\ee
where again we've used properties~(0.1) and (1.2) from the main text.
So
\be
\mathcal{C}^{(1)}_P(\mathcal{N})\leq\max_{\substack{\{q,P'\}\\ (1-q)P'=P}}q\mathcal{C}(\mathcal{N}_0)+(1-q)\mathcal{C}^{(1)}_{P'}(\mathcal{N}_1).
\ee
This upper bound can always be achieved, hence
\be
\mathcal{C}^{(1)}_P(\mathcal{N})=\max_{\substack{\{q,P'\}\\(1-q)P'=P}}q\mathcal{C}(\mathcal{N}_0)+(1-q)\mathcal{C}^{(1)}_{P'}(\mathcal{N}_1).
\ee

Essentially the same argument shows that
\be\label{eq:nshot}
n\mathcal{C}^{(n)}_P\left(\mathcal{N}\right)=\mathcal{C}^{(1)}_{nP}\left(\mathcal{N}^{\otimes n}\right)=\max_{\substack{\{q_k,P_k\}\\ \sum q_kP_k=nP}} \sum_{k=1}^{n}q_k\mathcal{C}^{(1)}_{P_k}\left(\mathcal{N}_0^{\otimes n-k}\otimes\mathcal{N}_1^{\otimes k}\right)+\left(1-\sum_{k=1}^nq_k\right)\mathcal{C}^{(1)}\left(\mathcal{N}_0^{\otimes n}\right).
\ee
Applying Lemma~\ref{lemma:CP_classicalquantum} (see below) to $ \mathcal{N}_0^{\otimes n-k} $ and $ \mathcal{N}_1^{\otimes k} $, we obtain
\be
\mathcal{C}^{(1)}_{P_k}\left(\mathcal{N}_0^{\otimes n-k}\otimes\mathcal{N}_1^{\otimes k}\right)=\mathcal{C}\left(\mathcal{N}_0^{\otimes n-k}\right)+\mathcal{C}^{(1)}_{P_k}\left(\mathcal{N}_1^{\otimes k}\right)=(n-k)\mathcal{C}\left(\mathcal{N}_0\right)+k\mathcal{C}^{(k)}_{P_k/k}\left(\mathcal{N}_1\right).
\ee
So Eq.~(\ref{eq:nshot}) becomes
\be
n\mathcal{C}^{(n)}_P\left(\mathcal{N}\right)=\max_{\substack{\{q_k,P_k\}\\\sum q_kP_k=nP}}\sum_{k=1}^{n}q_k\left((n-k)\mathcal{C}\left(\mathcal{N}_0\right)+k\mathcal{C}^{(k)}_{P_k/k}\left(\mathcal{N}_1\right)\right)+\left(1-\sum_{k=1}^nq_k\right)n\mathcal{C}\left(\mathcal{N}_0\right).
\ee
Since
\be
\mathcal{C}^{(k)}_{P}\left(\mathcal{N}_1\right)\leq \mathcal{C}_{P}\left(\mathcal{N}_1\right),
\ee
we have
\begin{align}
\mathcal{C}^{(n)}_P\left(\mathcal{N}\right)&\leq \max_{\substack{\{q_k,P_k\}\\\sum q_kP_k=nP}}\sum_{k=1}^{n}q_k\left(\frac{n-k}{n}\mathcal{C}\left(\mathcal{N}_0\right)+\frac{k}{n}\mathcal{C}_{P_k/k}\left(\mathcal{N}_1\right)\right)+\left(1-\sum_{k=1}^nq_k\right)\mathcal{C}\left(\mathcal{N}_0\right)\\
&\leq \max_{\{q_k\}}\left(1-\sum_{k=1}^n\frac{kq_k}{n}\right)\mathcal{C}\left(\mathcal{N}_0\right)+\left(\sum_{k=1}^n\frac{kq_k}{n}\right)\mathcal{C}_{P/\left(\sum_{k=1}^n\frac{kq_k}{n}\right)}\left(\mathcal{N}_1\right)\\
&=\max_{\substack{\{q,P'\}\\(1-q)P'=P}}q\mathcal{C}\left(\mathcal{N}_0\right)+(1-q)\mathcal{C}_{P'}\left(\mathcal{N}_1\right),
\end{align}
where in the second line, we used the concavity of $ \mathcal{C}_P $ with respect to $ P $. The third line is just a relabelling.
This implies
\be
\mathcal{C}_P\left(\mathcal{N}\right)\leq \max_{\substack{\{q,P'\}\\(1-q)P'=P}}q\mathcal{C}\left(\mathcal{N}_0\right)+(1-q)\mathcal{C}_{P'}\left(\mathcal{N}_1\right).
\ee
On the other hand, it is clear that	by taking $ q_k=0 $ for $ k<n $ in Eq.(\ref{eq:nshot}), we have
\be
\mathcal{C}^{(n)}_P\left(\mathcal{N}\right)\geq\max_{\substack{\{q,P'\}\\(1-q)P'=P}}q\mathcal{C}\left(\mathcal{N}_0\right)+(1-q)\mathcal{C}^{(n)}_{P'}\left(\mathcal{N}_1\right).
\ee
Taking the limit $ n\to \infty $ on both sides, we obtain the other direction of the inequality
\be
\mathcal{C}_P\left(\mathcal{N}\right)\geq \max_{\substack{\{q,P'\}\\(1-q)P'=P}}q\mathcal{C}\left(\mathcal{N}_0\right)+(1-q)\mathcal{C}_{P'}\left(\mathcal{N}_1\right).
\ee
Hence
\be
\mathcal{C}_{P}\left(\mathcal{N}\right)=\max_{\substack{\{q,P'\}\\(1-q)P'=P}}q\mathcal{C}\left(\mathcal{N}_0\right)+(1-q)\mathcal{C}_{P'}\left(\mathcal{N}_1\right).
\ee
\end{proof}
\subsection{Lemma~\ref{lemma:CP_classicalquantum}}
\begin{lemma}\label{lemma:CP_classicalquantum}
For classical channel $\Psi_0$ and quantum channel $\Psi_1$, the 1-shot classical capacity of $ \Psi_0\otimes \Psi_1 $ with limited entanglement assistance  satisfies
\be
\mathcal{C}^{(1)}_P\left(\Psi_0\otimes \Psi_1\right)=\mathcal{C}\left(\Psi_0\right)+\mathcal{C}^{(1)}_P\left(\Psi_1\right).
\ee
\end{lemma}
This lemma is very intuitive. Essentially it says the CE tradeoff region for a tensor product of classical and quantum channel is additive. Unfortunately we did not find any description of this result in the literature. So we provide our own proof here, using Lemma~\ref{lemma:cq} from the main text.\\

\begin{proof}
Since $ \Psi_0 $ is classical, there exists a classical noiseless channel $ \Pi $ such that $ \Psi_0=\Psi_0\otimes\left(\Pi\otimes \mathcal{I}_C\right) $. So $ \Psi_0\otimes\Psi_1 $ is a partial cq channel. Hence $ \mathcal{C}^{(1)}_P\left(\Psi_0\otimes\Psi_1\right) $ can be achieved with ensemble of the form $ \left\{p_{ij},\ket{j}\bra{j}\otimes\rho_{ij}\right\} $. For such an ensemble, $ \chi_{\rm assist}\left(\Psi_0\otimes\Psi_1,\left\{p_{ij},\ket{j}\bra{j}\otimes\rho_{ij}\right\}\right) $ is
\be
\sum_{ij}p_{ij}S\left(\rho_{ij}\right)+S\left(\Psi_0\otimes\Psi_1\left(\sum_{ij}p_{ij}\ket{j}\bra{j}\otimes\rho_{ij}\right)\right)-\sum_{ij}p_{ij}S\left(\Psi_0\otimes\Psi_1\otimes\mathcal{I}\left(\ket{j}\bra{j}\otimes\phi_{ij}\right)\right).
\ee
Notice that for the second term, by subadditivity of the von Neumann entropy,
\be
S\left(\Psi_0\otimes\Psi_1\left(\sum_{ij}p_{ij}\ket{j}\bra{j}\otimes\rho_{ij}\right)\right)\leq S\left(\Psi_0\left(\sum_{ij}p_{ij}\ket{j}\bra{j}\right)\right)+S\left(\Psi_1\left(\sum_{ij}p_{ij}\rho_{ij}\right)\right).
\ee
Also, since
\be
\Psi_0\otimes\Psi_1\otimes\mathcal{I}\left(\ket{j}\bra{j}\otimes\phi_{ij}\right)=\Psi_0\left(\ket{j}\bra{j}\right)\otimes\left(\Psi_1\otimes\mathcal{I}\right)\left(\phi_{ij}\right),
\ee
the third term is equivalent to
\be
\sum_{ij}p_{ij}S\left(\Psi_0\otimes\Psi_1\otimes\mathcal{I}\left(\ket{j}\bra{j}\otimes\phi_{ij}\right)\right)=\sum_{ij}p_{ij}S\left(\Psi_0\left(\ket{j}\bra{j}\right)\right)+\sum_{ij}p_{ij}S\left(\Psi_1\otimes\mathcal{I}\left(\phi_{ij}\right)\right).
\ee
So
\begin{align}
\mathcal{C}^{(1)}_P\left(\Psi_0\otimes\Psi_1\right)&\leq\max_{\substack{\{p_{ij},\rho_{ij}\}\\\sum_{ij}p_{ij}S\left(\rho_{ij}\right)\leq P}} S\left(\Psi_0\left(\sum_{ij}p_{ij}\ket{j}\bra{j}\right)\right)-\sum_{ij}p_{ij}S\left(\Psi_0\left(\ket{j}\bra{j}\right)\right)\nonumber\\
&+\sum_{ij}p_{ij}S\left(\rho_{ij}\right)+S\left(\Psi_1\left(\sum_{ij}p_{ij}\rho_{ij}\right)\right)-\sum_{ij}p_{ij}S\left(\Psi_1\otimes\mathcal{I}\left(\phi_{ij}\right)\right)\nonumber\\
&\leq \mathcal{C}\left(\Psi_0\right)+\mathcal{C}^{(1)}_{P}\left(\Psi_1\right).
\end{align}
On the other hand, by restricting input states to product states with respect to the two channels, it can be easily shown that
\be
\mathcal{C}^{(1)}_P\left(\Psi_0\otimes\Psi_1\right)\geq\mathcal{C}\left(\Psi_0\right)+\mathcal{C}^{(1)}_P\left(\Psi_1\right).
\ee
Hence
\be
\mathcal{C}^{(1)}_P\left(\Psi_0\otimes\Psi_1\right)=\mathcal{C}\left(\Psi_0\right)+\mathcal{C}^{(1)}_P\left(\Psi_1\right).
\ee
\end{proof}

\subsection{Construction of $ \mathcal{N}_0 $}
Let $ \mathcal{G}:B\to B $ be the classical symmetric channel of the form
\be
\mathcal{G}(\ket{k}\bra{k})=(1-\eta)\ket{k}\bra{k}+\eta I/|B|.
\ee
Its classical capacity is
\be
\mathcal{C}(\mathcal{G})=\log|B|-H(\eta),
\ee
where
\be
H(\eta)=H_2\left(\frac{|B|-1}{|B|}\eta\right)+\frac{|B|-1}{|B|}\eta\log\left(|B|-1\right)
\ee
and $ H_2 $ is the binary entropy.

Take $ \mathcal{H}_A $ with $ |A|>|B| $. Then $ \mathcal{H}_A$ has basis $ \{\ket{1},\dots,\ket{|B|},\ket{|B|+1}\dots,\ket{|A|}\} $. Define $ \mathcal{N}_0: A\to B $ as
\begin{align*}
\mathcal{N}_0(\ket{k}\bra{k})&=\mathcal{G}(\ket{k}\bra{k})&\text{if}&~~1\leq k\leq |B|,\\
\mathcal{N}_0(\ket{k}\bra{k})&=I/|B|&\text{if}&~~|B|+1 \leq k\leq |A|.
\end{align*}

Then $ \mathcal{N}_0 $ is a classical channel with the same classical capacity as $ \mathcal{G} $. It's clear that $ \mathcal{N}_0 $ satisfies properties~(0.1) and (0.2) from the main text.
\subsection{Properties of $ \mathcal{F} $}
We prove $ \mathcal{F}^{RC\to C} $ satisfies properties~(1.1) and (1.2) from the main text. 

It's clear that
\begin{align}
\mathcal{C}^{(1)}_P\left(\mathcal{F}\right)&=\max_{\substack{\{p_i,\rho_i\}\\ \sum_ip_iS\left(\rho_i\right)\leq P}}
\sum_ip_iS\left(\rho_i\right)+S\left(\mathcal{F}\left(\sum_ip_i\rho_i\right)\right)-\sum_ip_iS\left(\mathcal{F}\otimes\mathcal{I}\left(\phi_i\right)\right)\nonumber\\
&\leq \log(|C|)-\min_{S(\rho)\leq P}\left(S\left(\mathcal{F}\otimes\mathcal{I}\left(\phi\right)\right)-S\left(\rho\right)\right)
\end{align}
The above inequality is true for any channel $ \mathcal{F} $.

In proving Lemma~\ref{lemma:cq} from the main text, we basically show that with averaged input entropy constrained to be no more than $ P $, ensembles of the form $ \left\{p_{ij},\ket{j}\bra{j}\otimes\rho_{ij}\right\} $ minimizes $ \sum_ip_i\left(S\left(\mathcal{F}\otimes\mathcal{I}\left(\phi_{ij}\right)\right)-S\left(\rho_{ij}\right)\right) $.
\begin{align}
&\sum_{ij}p_{ij}\left(S\left(\mathcal{F}\otimes \mathcal{I}\left(\ket{j}\bra{j}\otimes\phi_{ij}\right)\right)-S\left(\rho_{ij}\right)\right)\nonumber\\
=&\sum_{ij}p_{ij}\left(S\left(X_j\mathcal{E}\otimes \mathcal{I}\left(\phi_{ij}\right)X_j^\dagger\right)-S\left(\rho_{ij}\right)\right)\nonumber\\
=&\sum_{ij}p_{ij}\left(S\left(\mathcal{E}\otimes\mathcal{I}\left(\phi_{ij}\right)\right)-S\left(\rho_{ij}\right)\right)
\end{align}

Suppose $ \rho $ minimizes $ S\left(\mathcal{E}\otimes\mathcal{I}\left(\phi\right)\right)-S\left(\rho\right) $ subject to $ S\left(\rho\right)\leq P $. By the above argument, it's clear that
\be
S\left(\mathcal{E}\otimes\mathcal{I}\left(\phi\right)\right)-S\left(\rho\right)=\min_{S(\rho')\leq P}S\left(\mathcal{F}\otimes\mathcal{I}\left(\phi'\right)\right)-S\left(\rho'\right)
\ee
Now consider the ensemble $ \left\{\frac{1}{|C|^2},\ket{j}\bra{j}\otimes\rho\right\} $
\begin{align}
&\chi_{\rm assist}\left(\mathcal{F},\left\{\frac{1}{|C|^2},\ket{j}\bra{j}\otimes\rho\right\}\right)\nonumber\\
=&S\left(\rho\right)+S\left(\sum_j\frac{1}{|C|^2}\mathcal{F}\left(\ket{j}\bra{j}\otimes\rho\right)\right)-\sum_j\frac{1}{|C|^2}S\left(\mathcal{F}\otimes \mathcal{I}\left(\ket{j}\bra{j}\otimes\phi\right)\right)\nonumber\\
=&S\left(\rho\right)+S\left(\sum_j\frac{1}{|C|^2}X_j^\dagger\mathcal{E}\left(\rho\right)X_j^\dagger\right)-\sum_j\frac{1}{|C|^2}S\left(X_j\mathcal{E}\otimes \mathcal{I}\left(\phi\right)X_j^\dagger\right)\nonumber\\
=&\log(|C|)+S\left(\rho\right)-S\left(\mathcal{E}\otimes\mathcal{I}\left(\phi\right)\right)
\end{align}
where we've used the following qudit twirl formula \cite{Wilde13_Sup}
\be
\frac{1}{|C|^2}\sum_jX_j\rho X_j^\dagger=\frac{I}{|C|}.
\ee
One can show a similar result for $ \mathcal{C}^{(n)}_P\left(\mathcal{F}\right) $. This shows $ \mathcal{F} $ satisfies property~(1.2) from the main text.

Let $ \phi $ be a state that achieves minimum output entropy for $ \mathcal{E} $. Consider the ensemble $ \ket{k}\bra{k}\otimes\phi $ with equal probability. The Holevo information of this ensemble is
\begin{align}
&\chi\left(\mathcal{F},\left\{\frac{1}{|C|^2},\ket{k}\bra{k}\otimes\phi\right\}\right)\nonumber\\
=&S\left(\frac{1}{|C|^2}X_k\mathcal{E}\left(\phi\right)X_k^\dagger\right)-\frac{1}{|C|}S\left(X_k\mathcal{E}\left(\phi\right)X_k^\dagger\right)\nonumber\\
=&\log(|C|)-S_{\text{min}}\left(\mathcal{E}\right)
\end{align}
For any ensemble $ \{p_{ij},\ket{j}\bra{j}^R\otimes\rho_{ij}^{C}\} $,
\be
S\left(\sum_{ij}p_{ij}\mathcal{F}\left(\ket{j}\bra{j}\otimes\rho_{ij}\right)\right)\leq \log(|C|)\\
\ee
and
\be
S\left(\mathcal{F}\left(\ket{j}\bra{j}\otimes\rho_{ij}\right)\right)=S\left(X_j\mathcal{E}\left(\rho_{ij}\right)X_j^\dagger\right)\geq S_{\text{min}}\left(\mathcal{E}\right).
\ee
Hence
\be
\mathcal{C}^{(1)}\left(\mathcal{F}\right)=\log(|C|)-S_{\text{min}}\left(\mathcal{E}\right).
\ee
Similarly it can be shown
\be
\mathcal{C}^{(n)}\left(\mathcal{F}\right)=\log(|C|)-S_{\text{min}}\left(\mathcal{E}^{\otimes n}\right)/n.
\ee
Since we've chosen $ \mathcal{E} $ with subadditive minimum output entropy, \textit{i.e.}
\be
S_{\text{min}}\left(\mathcal{E}\otimes\mathcal{E}\right)< 2S_{\text{min}}\left(\mathcal{E}\right),
\ee
the channel $ \mathcal{F} $ will satisfy property~(1.1) from the main text
\be
\mathcal{C}\left(\mathcal{F}\right)\geq \mathcal{C}^{(2)}\left(\mathcal{F}\right)>\mathcal{C}^{(1)}\left(\mathcal{F}\right).
\ee
	
\end{document}